# Magnetic imaging of antiferromagnetic domain walls

Paul M. Sass[1], Wenbo Ge[1], Jiaqiang Yan[2], D. Obeysekera[3], J.J Yang[3], and Weida Wu[1]*.

[1] *Department of Physics and Astronomy, Rutgers University, Piscataway, NJ 08854, USA.*

[2] *Materials Science and Technology Division, Oak Ridge National Laboratory, Oak Ridge, Tennessee 37831, USA.*

[3] *Department of Physics, New Jersey Institute of Technology, Newark, 07102 USA*

*Correspondence to: wdwu@physics.rutgers.edu (WW).

**Abstract:** The control of domain walls or spin textures is crucial for spintronic applications of antiferromagnets[1,2]. Despite many efforts, it has been challenging to directly visualize antiferromagnetic domains or domain walls with nanoscale resolution, especially in magnetic field. Here, we report magnetic imaging of domain walls in several uniaxial antiferromagnets, the topological insulator $MnBi_2Te_4$ family and the Dirac semimetal $EuMnBi_2$, using cryogenic magnetic force microscopy (MFM). Our MFM results reveal higher magnetic susceptibility or net moments inside the domain walls than in domains. Domain walls in these antiferromagnets form randomly with strong thermal and magnetic field dependences. The direct visualization of domain walls and domain structure in magnetic field will not only facilitate the exploration of intrinsic phenomena in topological antiferromagnets, but also open a new path toward control and manipulation of domain walls or spin textures in functional antiferromagnets.

Because of compensated magnetic moments, antiferromagnetic order produces no stray field, possesses fast dynamics, and is insensitive to magnetic perturbation[1]. Thus, antiferromagnets (AFMs) have been used as a pinning layer in spin-valve devices[1,2]. Furthermore, the robustness and non-invasiveness of AFMs make them appealing for replacing the ferromagnetic active components in spintronic devices[1–3]. In addition, recent progress in topological materials suggest that many AFMs may host interesting topological states[4]. For example, it has been proposed that an axion insulator state with topological magnetoelectric response could be realized in an antiferromagentic-topological insulator (AFM-TI) phase[5,6]. The AFM-TI state adiabatically connects to a stack of quantum Hall insulators with alternating Chern numbers[7], thus it also provides a promising route to realizing quantum anomalous Hall (QAH) effect in stoichiometric materials. The prior observation of the QAH effect in magnetically doped TI thin films is limited to extremely low temperature because of inherent disorders[8–12], though the disorder effect can be partially alleviated by material engineering[13–15]. The $MnBi_2Te_4$ family was predicted to be promising candidates of AFM-TI that may host QAH and axion insulator states in thin films with odd and even septuple layers (SLs), respectively[16–18]. Recent transport measurements on exfoliated thin flakes provide compelling evidence for these predictions[19,20]. Although there has been a surge of research efforts in this class of materials, there is no report on the domain structure in these materials. Multiple domains with opposite signs would cancel each other, resulting in vanishing topological magnetoelectric response or QAH effect[21]. Therefore, it is imperative to visualize and control AFM domains or domain walls (DWs) in these AFMs with topologically non-trivial band structure to explore topological phenomena[1,4,22]. The nanoscale imaging of AFM DWs would allow exploration of the chiral electronic states residing on edges or DWs[6,21].

Yet, it is technically challenging to visualize AFM domains or domain walls with magnetic probes due to their vanishing magnetization. Most AFM imaging techniques rely on the secondary effects induced by the AFM order, e.g. optical birefringence due to magnetostrictive or piezomagnetic effect, non-linear optical response due to broken inversion symmetry, or local electron spin polarization induced by AFM order. Examples of AFM imaging techniques include linear and nonlinear optical microscopy[23], x-ray magnetic linear dichroism with photo-emission electron microscopy [24–26], spin-polarized scanning tunneling microsscopy[27], and more recently, x-ray Bragg diffraction phase contrast microscopy [28]. However, none of them can visualize the domain evolution across the magnetic field induced spin-flop or spin-flip transition with nanoscale spatial resolution[29].

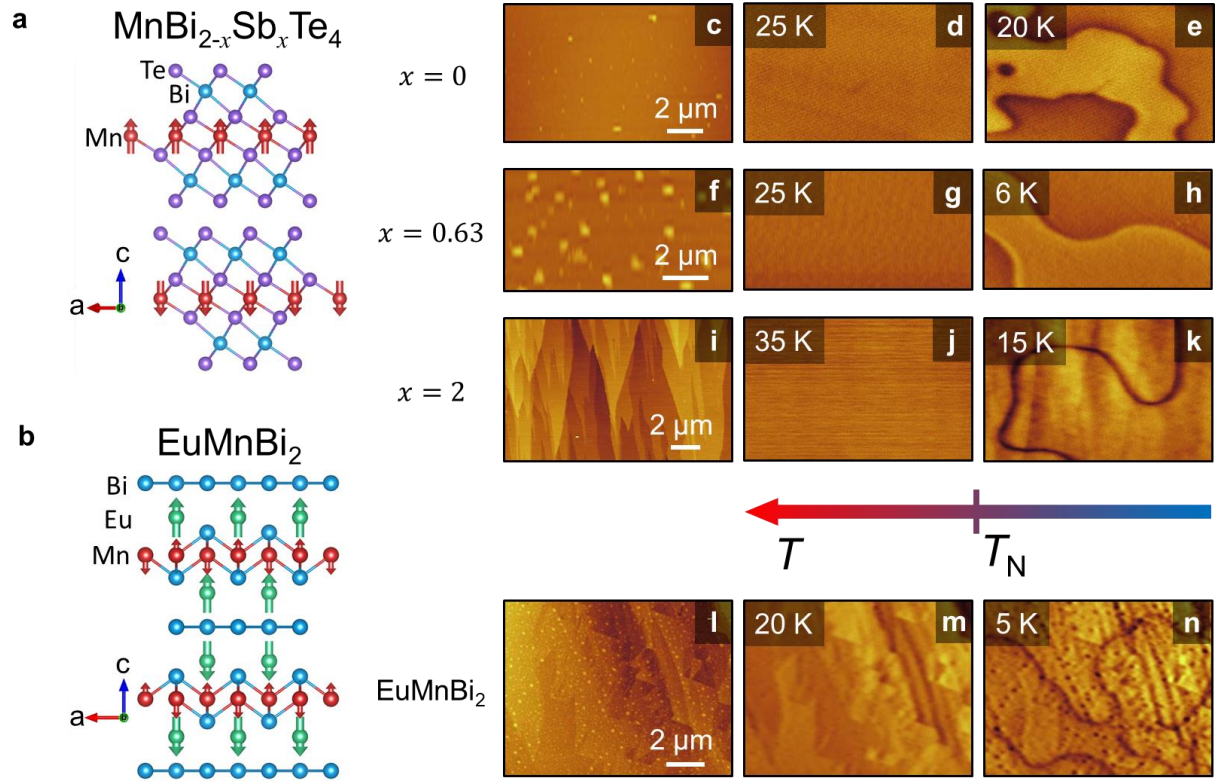

**Fig. 1 | Crystal structure, topographic and magnetic force microscopy (MFM) data of MnBi$_{2-x}$Sb$_x$Te$_4$ and EuMnBi$_2$. a,b,** A schematic illustration of crystal structure and magnetic order of Mn(Bi,Sb$_2$)Te$_4$ and EuMnBi$_2$, respectively. Note that in EuMnBi$_2$, only Eu moments form A-type AFM order. **c,f,i,l,** Topographic images of MnBi$_{2-x}$Sb$_x$Te$_4$ (MBST) singe crystals ($x = 0, 0.63, 2$) and EuMnBi$_2$, respectively. The color scales are: 25, 50, 15 and 40 nm, respectively. **d,g,j,m,** MFM images taken on these samples above $T_N$. No domain wall (DW) is visible. **e,h,k,n,** MFM images taken on these samples below $T_N$ at zero field (MBST) and 5.0 T (EuMnBi$_2$). The color scales are: 0.15, 0.3, 0.2 and 1.0 Hz, respectively. The Néel ordering temperatures ($T_N$) are 24, 23, 19, and 25 K for MBST ($x = 0, 0.63, 2$) and EuMnBi$_2$, respectively.

Here, we report direct visualization of AFM DW in single crystals of MnBi$_{2-x}$Sb$_x$Te$_4$ (MBST, $x = 0, 0.63, 1$) and Dirac semimetal EuMnBi$_2$ using cryogenic magnetic force microscopy (MFM) with *in-situ* transport. (See supplementary section 1 for transport data and *H-T* phase diagrams) All three MBST members are uniaxial AFM with uniaxial A-type AFM order, i.e., alternating ferromagnetic Mn order in SLs (Fig. 1**a**)[30], while in EuMnBi$_2$ only the Eu moments form A-type AFM order (Fig. 1**b**)[31]. Quantum-Hall-like transport was observed in topological Dirac semimetal EuMnBi$_2$ above the spin-flop transition of the Eu moments[31]. The $x = 0.63$ MBST is particularly appealing for exploring axion insulator physics, because it is close to the crossover between *n*-type (MnBi$_2$Te$_4$) and *p*-type (MnSb$_2$Te$_4$) conduction[30]. Our MFM images reveal that DWs of A-type AFM order emerge below the Néel ordering temperature ($T_N \sim$ 19-25 K) in these topological antiferromagnets. Field dependence of DW signals suggest that the magnetic contrast of DWs comes from the enhanced susceptibility or net magnetization inside DWs. This is caused by the winding of AFM order parameter across a DW, resulting in a spin-flop or spin-flip state inside the DW. This contrast mechanism might be generalized to imaging DWs in AFMs with uniaxial anisotropy. The observed DW width is ~500 nm, which is possibly resolution limited. The domain configuration is randomized after a thermal cycle to $T > T_N$ or a field cycle to the saturated state, indicating random nucleation and weak pinning.

For an A-type AFM with uniaxial anisotropy, there are only two possible domain states, up-down-up-down (↑↓↑↓) and down-up-down-up (↓↑↓↑)[32]. They are related to each other by either time reversal symmetry or a fractional lattice translation, so they are antiphase domains and the AFM DWs separating them are antiphase boundaries. Therefore, there won't be any vertex point connected to three or more DWs. Indeed, this is what we observed in all four AFMs. As shown in Fig. 1, curvilinear features with dark contrast emerge only below $T_N$. These features are either continuous or forming loops without any vertex point or preferred orientation, which are consistent with only two antiphase domain states in these systems. Thus, we conclude that these curvilinear features are antiphase AFM DWs. The observation of DWs in these uniaxial AFMs with very different crystal structure and magnetic couplings suggest that magnetic imaging of AFM DWs can be generalized to other functional AFMs.

Note that the AFM DWs in MnSb$_2$Te$_4$ ($x = 2$) sample show dark contrast at zero field, indicating that they carry finite magnetic moments parallel to the MFM tip moment. In contrast, domain contrast between two AFM domain states was observed in MBST (*x*=0 and 0.63) at zero fields. The metamagnetic transition is of spin-flip type ($H_{SF}$~0.3 T) in MnSb$_2$Te$_4$, suggesting the interlayer AFM exchange coupling is much weaker than anisotropy energy[30]. This energetics is similar to that in synthetic AFM (effectively A-type) widely used in spintronics[2,3,33], indicating a similar ferromagnetic configuration inside DWs. However, such ferromagnetic DW has not been reported in natural A-type AFMs[34]. We note that AFM DWs with net moment has been observed in the multiferroic domain boundaries of Z$_6$ vortex domains in hexagonal manganites, where Mn spins form 120° spin order[35]. These AFM DWs are pinned to the structural antiphase-ferroelectric domain walls because of cross-coupling[36]. Note that weak domain contrast due to imperfect compensation near surface was also reported in synthetic AFM[37,38]. Zero field domain contrast is invisible in MnSb$_2$Te$_4$, but is clearly observed in MBST ($x = 0$ and 0.63). This is likely due to much smaller saturation moment (~1.5 μ$_B$/Mn) of MnSb$_2$Te$_4$ than that of MBST (~3.2-3.5 μ$_B$/Mn)[30].

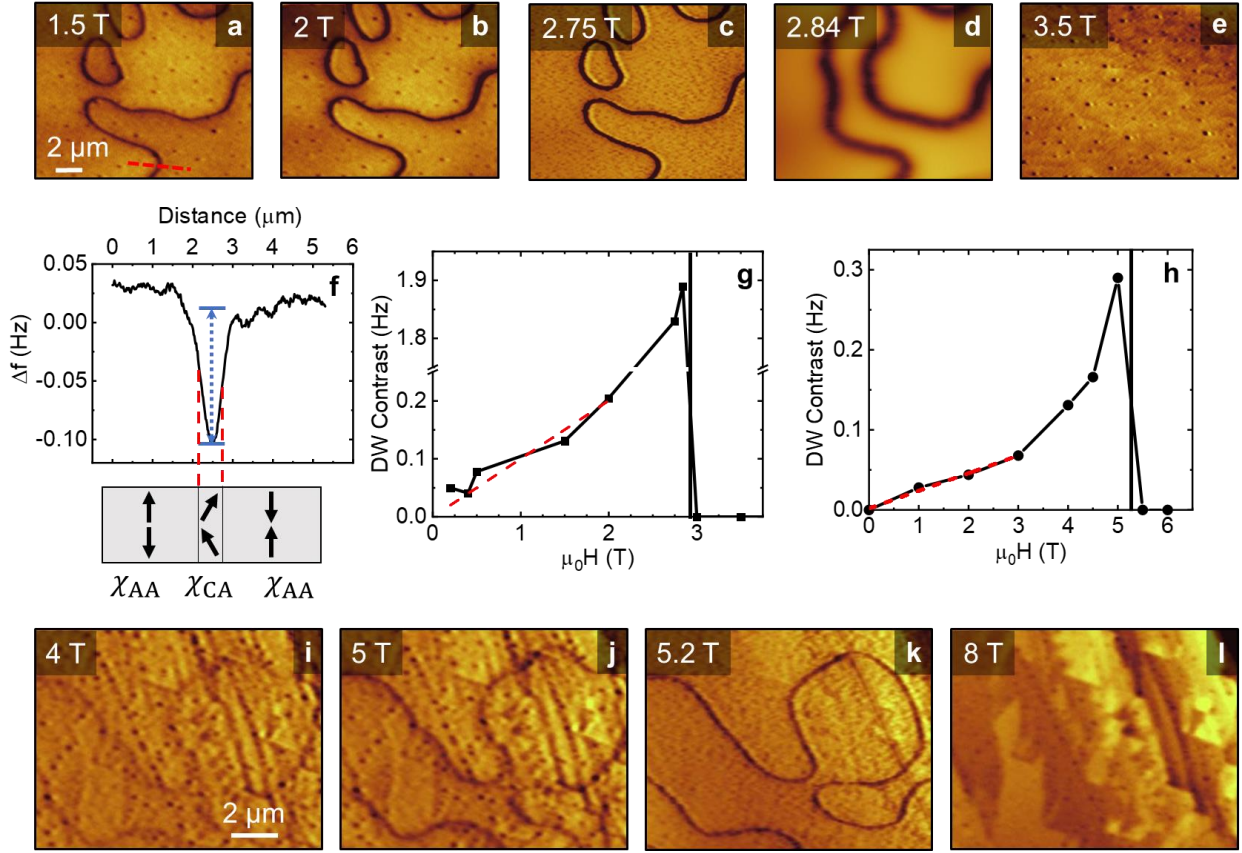

**Fig. 2 | MFM images and field dependence of DW contrast of MnBi$_{1.37}$Sb$_{0.63}$Te$_4$ and EuMnBi$_2$. a-d,** MFM images taken after 0.5 T FC on the same location of MnBi$_{1.37}$Sb$_{0.63}$Te$_4$ with increasing magnetic fields of 1.5, 2.0, 2.75, 2.84 T (labeled on upper left corner). Panel **c** and **d** show the coalescence of AFM domains. **e,** MFM image (3.5 T) of the canted AFM state. **f,** Line profile of DW (1.5 T) and schematic of spin structure of DW. The DW contrast (indicated by blue dash arrow) comes from the susceptibility difference between domains ($\chi_{AA}$) and DW ($\chi_{CA}$). **g,h,** $H$-dependence of DW contrast of MnBi$_{1.37}$Sb$_{0.63}$Te$_4$ and EuMnBi$_2$, respectively. It is approximately linear for $H < 2$ T, suggesting a susceptibility mechanism. The red dash line is a linear fit of the low field data. **i-l,** MFM images taken on the same location of EuMnBi$_2$ with increasing magnetic fields of 4.0, 5.0, 5.2, and 8.0 T (labeled on upper left corner).

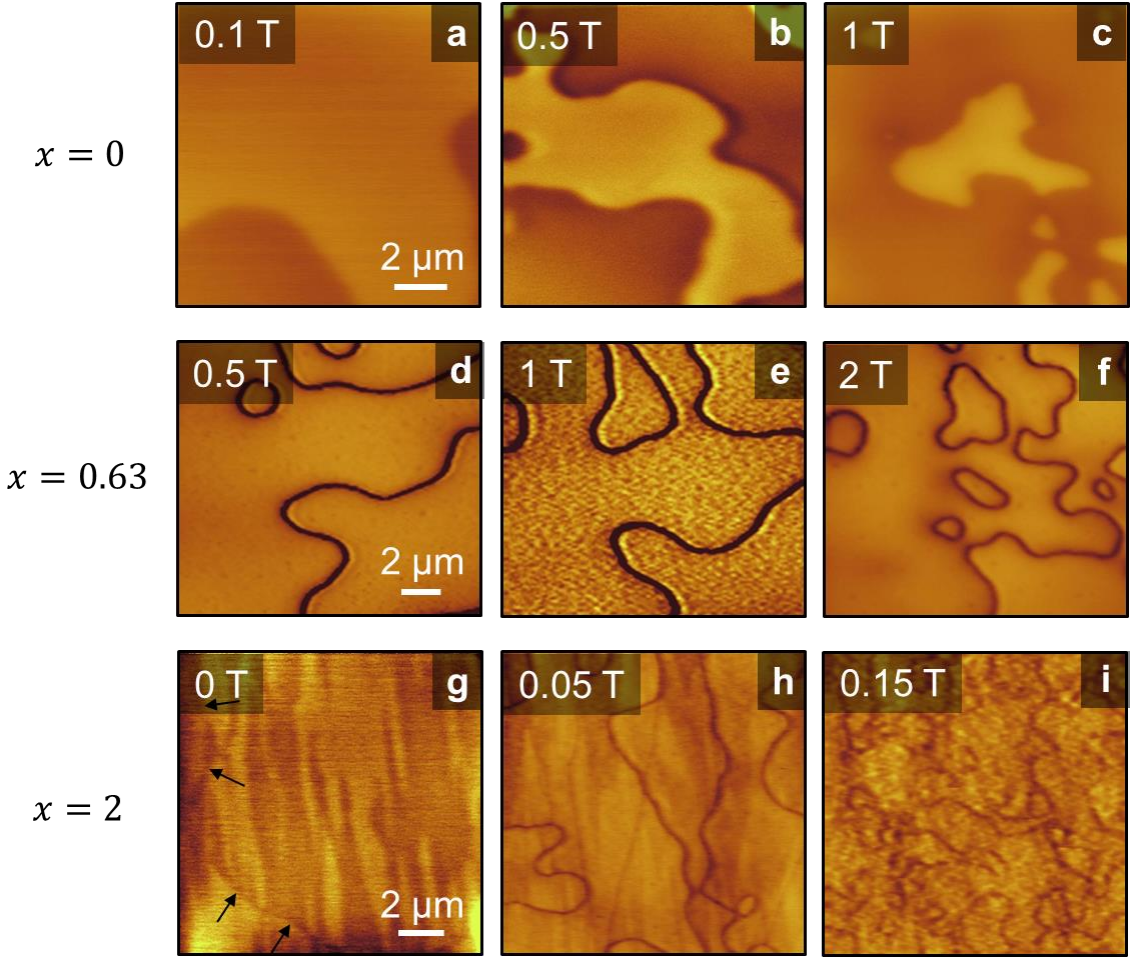

**Fig. 3 | Field cooled domain states of MnBi$_{2-x}$Sb$_x$Te$_4$ crystals. a-c**, MFM images at 5 K of $x$=0 (MBT) after 0.1, 0.5 and 1.0 T field cooling. **d-f**, MFM images at 5 K of $x$=0.63 (MBST) after 0.5, 1.0 and 2.0 T field cooling. **g-i**, MFM images at 5 K of $x$=2 (MST) after 0.1, 0.5 and 1.0 T field cooling. The arrows in panel **g** indicate the location of a DW. Cooling through $T_N$ with higher field induces more AFM DWs.

In contrast to MnSb$_2$Te$_4$, the metamagnetic transition ($H_{SF}$ ~3-3.5 T) of MBST ($x = 0$ and 0.63) is a spin-flop one, i.e., from the A-type AFM state to the canted AFM (CAFM) state[30]. This suggests a significantly stronger interlayer AFM exchange coupling[30]. The spin-flop transition is followed by a saturation transition ($H_S$ ~7-8 T) from the CAFM state to the forced ferromagnetic state[30]. These transitions are higher ($H_{SF}$~5.3 T and $H_S$~20 T) in EuMnBi$_2$, presumably enhanced by the *d-f* coupling between the Eu and Mn orders[31]. Therefore, it is of fundamental interests to explore the evolution of antiphase AFM DWs in these systems in high magnetic field. As shown in Figure 2**a**-2**d** and 2**i**-2**l**, the DW contrast of MBST (0.63) and EuMnBi$_2$ are substantially enhanced in finite magnetic fields. Similar behavior is also observed in MnBi$_2$Te$_4$. (See supplementary section 2) Because the AFM order must rotate away from its easy axis inside the DW, the spins of the antiphase DWs in MBST ($x$=0 and 0.63) and EuMnBi$_2$ are likely in the CAFM state as illustrated in Fig. 2**f**. Thus, the magnetic susceptibility of the DW is different from that of domain, which can be used for susceptibility imaging in high magnetic field[39]. Indeed, the DW contrast of MBST increases linearly with increasing magnetic field, then rises sharply right before the spin-flop transition as shown in Fig. 2**g** and 2**h**. These results confirm that the DW contrast in MBST and EuMnBi$_2$ originates from the susceptibility difference between the A-type AFM ($\chi_{AA}$) and the CAFM ($\chi_{CA}$) states[40]. (see supplementary section 2 for complete data sets) Therefore, our MFM results demonstrate that it is possible to visualize AFM DW in high magnetic field using the susceptibility contrast mechanism.

The observed DW width is ~500 nm, which is probably limited by spatial resolution because it is much larger than the estimated value ($\lesssim$ 10 nm) from exchange and anisotropy energy[40]. (See supplementary section 3) The sharp rise of DW contrast near the spin-flop transition indicates an increase of DW width, *i.e.* the volume of CAFM state (Fig. 2**d**). The DWs in MBST and EuMnBi$_2$ disappear above the spin-flop transition (Fig. 2**e** and 2**l**), suggesting a single domain state within the field of view. In the CAFM state, one expects 3 additional orientation variants (besides the 2 antiphase variants) due to trigonal crystallographic symmetry. CAFM domains in Cr$_2$O$_3$ has been visualized by non-linear optics[29]. Thus, one would expect similar multi-domain state in the CAFM states of MBST and EuMnBi$_2$. The observed single (large) domain state indicates very few nucleation sites for CAFM domain states in MnBi$_2$Te$_4$, MBST (0.63) and EuMnBi$_2$.

Because DW energy is proportional to the geometric mean of exchange and anisotropic energies, the larger anisotropy of MBST suggests an enhanced DW energy, thus favors less DW density. On the other hand, the Zeeman energy gain in high magnetic field lowers the DW energy, thus favors higher DW density. Although AFM domains are insensitive to small magnetic field because of cancellation of magnetization, DWs with significant net moments or higher magnetic susceptibility would be energetically more favorable in high magnetic field. Indeed, this simple scenario is confirmed by our MFM results. As shown Figure 3, higher DW density was observed after field cooling through $T_N$ with higher magnetic field values in all three MBST samples.

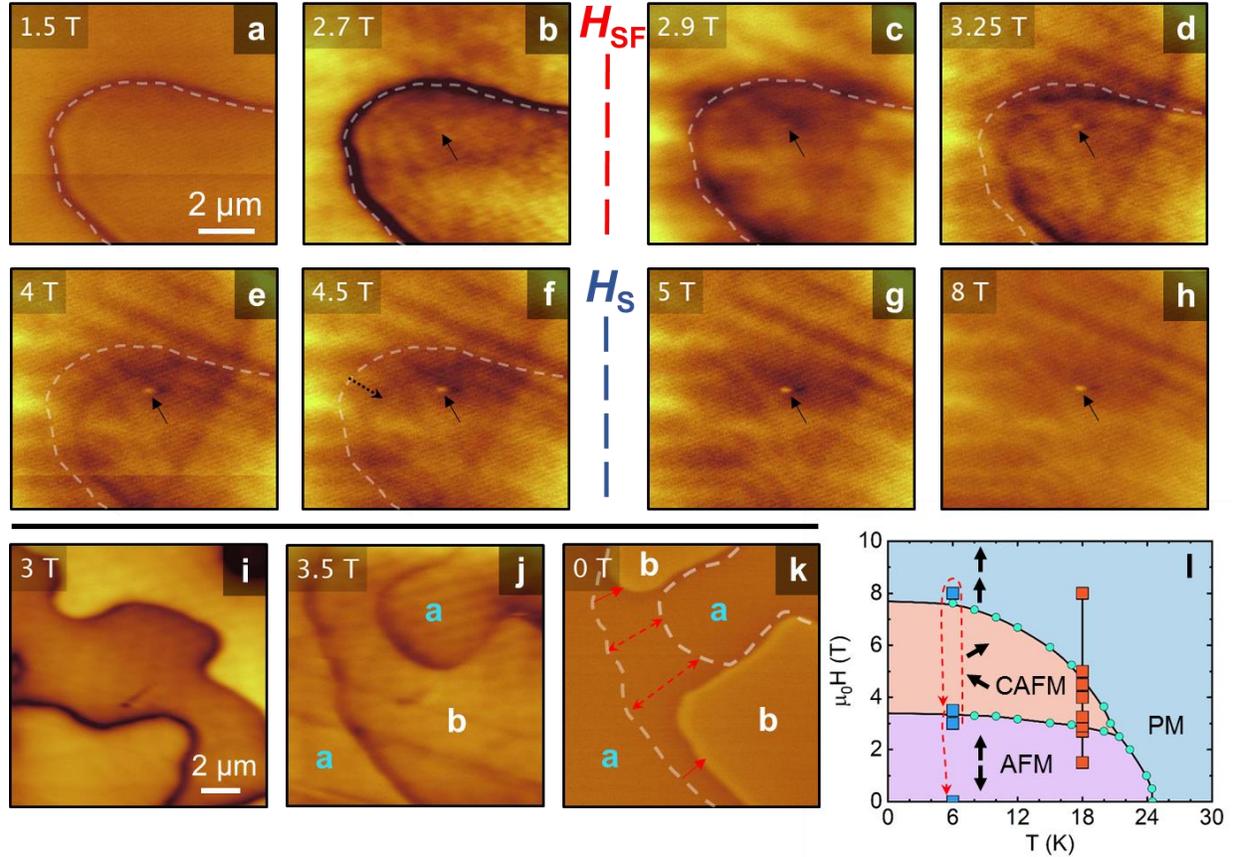

**Fig. 4 | *H*-dependence of DW and *H-T* phase diagram of MnBi$_2$Te$_4$.** **a-h**, MFM images at 18 K with increasing magnetic field of a MnBi$_2$Te$_4$ single crystal showing a robust AFM DW. The color scales are 0.5 and 0.2 Hz for **a-c** and **d-h**, respectively. The *H* values are labelled with orange squares in the *H-T* phase diagram in panel **l**. The $H_{SF}$ and $H_S$ represent the spin-flop and saturation transitions, respectively. A white dotted line outlines the DW observed at 1.5 T. The small solid black arrows note a topographic feature used for alignment. **i-k**, Field cycle on a different location. The color scale is 1 Hz. **i**, AFM domain pattern (3 T) on the upward ramping. **j**, At 3.5 T, on downward ramping, a different AFM domain pattern nucleates upon reentry into CAFM state; **k**, At 0 T, DWs creep and partial annihilation across $H_{SF}$. The white dash lines in **k** denote the DW locations at 3.5 T. The red dotted arrows indicate the partial annihilation of DWs, and the solid red arrows indicates the creep of DWs. **l**, *H-T* phase diagram. The orange and blue squares show the field values of MFM measurements.

In contrast to those in MnSb$_2$Te$_4$, MBST (0.63), and EuMnBi$_2$, the DWs in MnBi$_2$Te$_4$ survive above the spin-flop transition ($H_{SF}$), then disappear above the saturation transition ($H_S$). Figures 4**a**-4**h** show representative MFM images taken with upward ramping of magnetic field at 18 K ($H_{SF} \sim 2.87$ T and $H_S \sim 4.7$ T). The DW creeps a bit with increasing magnetic field, again indicating very weak pinning. The persistence of antiphase DWs across the spin-flop transition indicates a coherent rotation of all spins, including those in DWs. The coherent rotation of spins is probably induced by a small in-plane magnetic field component due to a slight sample tilt (~5°), which breaks the three-fold in-plane symmetry. Yet, the persistence of DWs in the CAFM state is perplexing because there is no obvious symmetry or other constraint to enforce coherent spin rotation inside the DWs. Such behavior might be explained if the exact spin structure of the DWs and its evolution across the spin-flop transition is resolved, e.g. by spin polarized scanning tunneling microscopy[27], which is beyond the scope of this work. No DW is observed above the saturation transition, consistent with the saturation state where all spins are aligned with magnetic field (Fig. 4**h**). Interestingly, a new multi-domain state (size ~10 µm) with DWs reemerges in the CAFM state (Fig. 4**i**-4**j**) after reducing magnetic field from 8 T. Here the two variants of AFM domains are labelled by **a** and **b**. Further reducing magnetic field causes creep and annihilation of DWs. At 0 T, the two DWs partially annihilates with each other, resulting in a different AFM domain pattern. The DW annihilation confirms that there are only two variants of AFM states above the spin-flop transition within the field of view. (See supplementary section 4 for more data on DW creep and annihilation)

Our results demonstrate that DWs of A-type AFM order in MBST ($x = 0, 0.63$ and 2) and EuMnBi$_2$ and their evolutions in magnetic field can be visualized by MFM utilizing the susceptibility (magnetization) difference between DWs and domains[39]. This susceptibility contrast mechanism is different from the ferromagnetic core mechanism established in synthetic AFM[34], which explains the DW contrast in MST. For DWs to be in the flop state, the only requirement is uniaxial anisotropy, which is quite general and is satisfied in many functional AFMs[1]. Recent advances in spintronics and 2D materials reveal exciting properties in uniaxial AFMs, e.g. spin Seebeck effect in MnF$_2$ heterostructures[41], giant tunneling magnetoresistance using CrI$_3$ flakes[42,43], and quantum transport in Dirac semimetal EuMnBi$_2$ and related compounds[44]. The visualization and manipulation of DWs in these materials will help to understand the fundamental mechanisms of these fascinating phenomena and their potential applications. Similarly, imaging and control of DWs in antiferromagnetic topological insulators such as MnBi$_{2-x}$Sb$_x$Te$_4$ will facilitate the exploration of chiral edge states at the DWs[6], and the realization of a single domain state, which is necessary for an unambiguous observation of the axion insulator and QAH states[19–21]. The weakly pinned DWs observed in MBST family and EuMnBi$_2$ might be manipulated by electric current via spin-transfer-torque[33,45], which could lead to low power logic or memory devices.

## Methods

**Sample preparation**

MnBi$_{2-x}$Sb$_x$Te$_4$: Platelike single crystals were grown out of a Bi(Sb)-Te flux and have been well characterized by measuring the magnetic and transport properties. All three compositions order magnetically below $T_N$ = 19-24 K with ferromagnetic Mn-Te layers coupled antiferromagnetically. MnBi$_{2-x}$Sb$_x$Te$_4$ with $x = 0.63$ was investigated in this work because it stays close to the transition from an n-type to p-type conducting behavior. At 2 K, MnSb$_2$Te$_4$ shows a spin flip transition at ~ 0.3 T with a magnetic field applied along the crystallographic c-axis. In contrast, MnBi$_2$Te$_4$ (MnBi$_{1.37}$Sb$_{0.63}$Te$_4$) shows a spin-flop transition at $H_{SF}$ =3.5 T (3.0 T) followed by moment saturation at $H_S$ = 7.8 T (6.8 T).

EuMnBi$_2$: single crystals were grown using Bi self-flux method using high purity Eu, Mn and Bi in 1:1:9 ratio placed in alumina crucibles in an Argon filled glove box and sealed in evacuated quartz tubes heated at 1000 ºC for 10 hours, followed by cooling to 400 ºC at 2 ºC/hour. The excess Bi flux was removed by centrifuging to obtain plate like single crystals.

**MFM measurement** The MFM experiments were carried out in a homemade cryogenic magnetic force microscope using commercial piezoresistive cantilevers (spring constant $k \approx$ 3 N/m, resonant frequency $f_0 \approx$ 42 kHz). The homemade MFM is interfaced with a Nanonis SPM Controller (SPECS) and a commercial phase-lock loop (SPECS). MFM tips were prepared by depositing nominally 150 nm Co film onto bare tips using e-beam evaporation. MFM images were taken in a constant height mode with the scanning plane nominally ~100 nm (except specified) above the sample surface. The MFM signal, the change of cantilever resonant frequency, is proportional to out-of-plane stray field gradient. Electrostatic interaction was minimized by nulling the tip-surface contact potential difference. Dark (bright) regions in MFM images represent attractive (repulsive) magnetization, where magnetizations are parallel (anti-parallel) with the positive external field.

***In-situ* transport measurement**     The Hall and longitudinal resistances were measured with standard lock-in technique with ac current of 0.1 – 4 mA modulated at 314 Hz.

**Data availability**

All data needed to evaluate the conclusions in the paper is available in the main text or the supplementary materials. Additional data requests should be addressed to the corresponding author.

**Acknowledgments**

The work at Rutgers is supported by DOE grant DE-SC0018153. Work at ORNL and Ames Laboratory was supported by the US Department of Energy, Office of Science, Basic Energy Sciences, Materials Sciences and Engineering Division.

**Author contributions**

W.W. conceived and supervised the project. J.Y. synthesized the single crystals of MBST family and performed characterization. J.J.Y. synthesized the single crystals of EuMnBi$_2$ and D.O. performed characterization. P.S. and W.G. carried out the MFM and *in-situ* transport measurements, and analyzed the results with W.W.'s supervision. P.S. and W.W. wrote the manuscript with contributions from all authors.

**Competing financial interests**

Authors declare no competing financial interests.